\documentclass[final]{cimento}
\input psfig.tex
\def\lsim{\lower.5ex\hbox{$\; \buildrel < \over \sim \;$}}
\def\gsim{\lower.5ex\hbox{$\; \buildrel > \over \sim \;$}}
\def\ch{\lower-0.55ex\hbox{--}\kern-0.55em{\lower0.15ex\hbox{$h$}}}
\def\lh{\lower-0.55ex\hbox{--}\kern-0.55em{\lower0.15ex\hbox{$\lambda$}}}
\title{Dirac equation in Kerr geometry and its solution}
\author{Sandip~K.~Chakrabarti\from{ins:x} and Banibrata Mukhopadhyay\from{ins:x}}
\instlist{\inst{ins:x} S. N. Bose National Centre for Basic Sciences,
JD-Block, Sector III, Salt Lake, Calcutta 700091, India
emails: chakraba@boson.bose.res.in and bm@boson.bose.res.in}
\PACSes{
\PACSit{04.20.-q}{Classical general relativity}
\PACSit{04.70.-s}{Physics of black holes}
\PACSit{04.70.Dy}{Quantum aspects of black holes}
\PACSit{95.30.Sf}{Relativity and gravitation}}
\begin{document}

\maketitle

\begin{abstract}

Chandrasekhar separated the Dirac equation for spinning and massive particles
in Kerr geometry into radial and angular parts. In the present review, we present solutions of 
the complete wave equation and discuss how the Dirac wave scatters off Kerr black holes.
The eigenfunctions, eigenvalues and reflection and transmission co-efficients
are computed for different Kerr parameters. We compare the solutions with several parameters to
show how a spinning black hole distinguishes mass and energy of incoming waves.
Very close to the horizon, the solutions become independent of the particle parameters
indicating an universality of the behaviour.

\end{abstract}

\noindent {Il Nuovo Cimento (in press)}

\section{Introduction}
One of the most important solutions of Einstein's equation is that
of the spacetime around and inside an isolated black hole.
The spacetime at a large distance is flat and Minkowskian
where usual quantum mechanics is applicable, while the spacetime
closer to the singularity is so curved that no satisfactory
quantum field theory could be developed as yet. An intermediate
situation arises when a weak perturbation (due to, say, gravitational,
electromagnetic or Dirac waves) originating from
infinity interacts with a black hole.
The resulting wave is partially transmitted into the
black hole through the horizon and partially scatters off
to infinity. In the linearized (`test field') approximation
this problem has been attacked in the past by several authors \cite{ref:t72, ref:t73, ref:c76, ref:c83}.
%(Teukolsky 1972; Teukolsky 1973; Chandrasekhar 1976; Chandrasekhar 1983)
The master equations of Teukolsky \cite{ref:t73} which govern these linear
perturbations for integral spin (e.g., gravitational and electromagnetic)
fields were solved numerically by Press \& Teukolsky \cite{ref:pt73}  and
Teukolsky \& Press \cite{ref:tp74}. While the equations governing the massive
Dirac particles were separated in 1976 \cite{ref:c76}, the
angular eigenfunction and eigenvalue (which happens to be the separation
constant) have been obtained in 1984 \cite{ref:c84} and radial solutions have been 
obtained only recently \cite{ref:bmskc98, ref:bmskc99, ref:skcbm00, ref:bm00}.
Particularly interesting is the fact that
whereas gravitational and electromagnetic radiations were found
to be amplified in some range of incoming frequencies,
Chandrasekhar \cite{ref:c83} predicted that no such amplifications should
take place for Dirac waves because of the very nature of the
potential experienced by the incoming fields. Although 
this later conclusion was drawn using an asymptotic equation,
we show that this is indeed the case even when complete solutions are
considered for the Dirac wave perturbations. Chandrasekhar also 
speculated that one needs to look into the problem for negative 
eigenvalues ($\lh$) where one {\it might} come across super-radiance 
for Dirac waves.

In the present review, we discuss this important problem and its 
solutions. We show the
nature of the radial wave functions as a function of the
Kerr parameter, rest mass and frequency of incoming particle.
We also verify that super-radiance is indeed absent for the Dirac field.
Unlike earlier works \cite{ref:pt73, ref:tp74} where numerical (shooting) methods were used to
solve the master equations governing
gravitational and electromagnetic waves, we use a classical approach whereby we approximate the
potential felt by the particle by  a collection of small steps.
 
Below, we present the separated Dirac equations
from Chandrasekhar \cite{ref:c83} using the same choice of units: we choose $\ch=1=G=c$,
so that the unit of mass becomes $\sqrt{\frac{\ch c}{G}}$, 
the unit of time becomes $\sqrt{\frac{\ch G}{c^5}}$, and the unit 
of length becomes $\sqrt{\frac{\ch G}{c^3}}$.

The equations governing the radial wave-functions $R_{\pm \frac{1}{2}}$
corresponding to spin $\pm \frac{1}{2}$ respectively are given by:
$$
\Delta^{\frac{1}{2}}{\cal D}_{0} R_{- \frac{1}{2}}
= ( \lh + i m_p r) \Delta^{\frac{1}{2}} {R}_{+ \frac{1}{2}} ,
\eqno{(1a)}
$$
$$
\Delta^{\frac{1}{2}} {\cal D}_{0}^{\dag} \Delta^{1 \over 2}
{R}_{+{\frac{1}{2}}} = ( \lh - i m_p  r)  {R}_{-{1 \over 2}} ,
\eqno{(1b)}
$$
where, the operators ${\cal D}_n$ and ${\cal D}_{n}^{\dag}$ are given by,
$$
{\cal D}_n = \partial_{r} + \frac {i K} {\Delta} + 2n \frac{(r-M)} {\Delta} ,
\eqno{(2a)}
$$
$$
{\cal D}_n^{\dag} = \partial_{r} - \frac {i K} {\Delta} + 2n \frac{(r-M)} {\Delta} ,
\eqno{(2b)}
$$
and
$$
\Delta = r^2 + a^2 - 2Mr ,
\eqno{(3a)}
$$
$$
K=(r^2 + a^2)\sigma + am .
\eqno{(3b)}
$$
Here, $a$ is the Kerr parameter, $n$ is an integer or half integer, $\sigma$ is the
frequency of incident wave, $M$ is the mass of the black hole, 
$m_p$ is the rest mass of
the Dirac particle, $\lh$ is the eigenvalue of the Dirac equation and 
$m$ is the azimuthal quantum number.

The equations governing the angular wave-functions $S_{\pm \frac{1}{2}}$ corresponding
to spin $\pm \frac{1}{2}$ respectively are given by:
$$
{\cal L}_{1 \over 2} S_{+ {1 \over 2}} = - (\lh -a m_p \cos \theta) S_{- {1 \over 2}}
\eqno{(4a)}
$$

$$
{\cal L}_{1 \over 2}^{\dag} S_{- {1 \over 2}} = + (\lh +a m_p \cos \theta) 
S_{+ {1 \over 2}}
\eqno{(4b)}
$$
where, the operators ${\cal L}_n$ and ${\cal L}_{n}^{\dag}$ are given by,

$$
{\cal L}_n = \partial_\theta + Q + n \cot \theta ,
\eqno{(5a)}
$$

$$
{\cal L}_n^{\dag} = \partial_\theta - Q + n \cot \theta
\eqno{(5b)}
$$
and
$$
Q=a \sigma \sin \theta + m \  {\rm cosec} \  \theta .
\eqno{(6)}
$$

Note that both the radial and the angular sets of equations i.e., 
eqs. 1(a-b)  and eqs. 4(a-b) are coupled equations. Combining eqs. 
4(a-b), one obtains the angular eigenvalue equations for the
spin-$\frac{1}{2}$ particles as \cite{ref:c84}
$$
\left [{\cal L}_{1 \over 2} {\cal L}_{1 \over 2}^{\dag} +
\frac{a m_p \sin \theta}{\lh + a m_p \cos \theta} {\cal L}_{1\over 2}^{\dag} +
(\lh^2 - a^2 m_p^2 \cos^2 \theta) \right ] S_{-{1 \over 2}} = 0 .
\eqno{(7)}
$$

There are exact solutions of this equation for the eigenvalues 
$\lh$ and the eigenfunctions
$S_{-{1\over 2}}$ when $\rho=\frac{m_p}{\sigma}=1$ 
in terms of the orbital quantum number $l$
and azimuthal quantum number $m$. These solutions are  \cite{ref:c84}:
$$
\lh^2 = (l+\frac{1}{2})^2 + a \sigma ( p+ 2m) + 
a^2 \sigma^2 \left [1-\frac{y^2}{2(l+1)+a\sigma x} \right ] ,
\eqno{(8)}
$$
and
$$
{}_{1\over 2}S_{lm} =
{}_{1\over 2}Y_{lm} - \frac{a\sigma y}{2(l+1)+a\sigma x} {}_{1\over 2}Y_{l+1 m}
\eqno{(9)}
$$
where,
$$
p=F(l,l); \ \ \ x=F(l+1,l+1); \ \ \ y=F(l,l+1)
$$
and
$$
F(l_1,l_2)=[(2l_2+1)(2l_1+1)]^{\frac{1}{2}} <l_2 1 m 0|l_1 m>
$$
$$
[<l_2 1 \frac{1}{2} 0|l_1 \frac{1}{2}> +(-1)^{l_2-l}<l_2 1 m 0|l_1 m>
[<l_2 1 \frac{1}{2} 0|l_1 \frac{1}{2}> +(-1)^{l_2-l}
\rho \sqrt{2} <l_2 1 -\frac{1}{2} 1|l_1 \frac{1}{2}>] .
\eqno{(10)}
$$
with $<....|..>$ are the usual Clebsh-Gordon coefficients.
For other values of $\rho$ one has to use perturbation theories. 
Solutions upto sixth order
using perturbation parameter $a\sigma$ are given in Chakrabarti \cite{ref:c84} and are not
described here. The eigenfunctions $\lh$ are required to solve the radial equations which we do now.

The radial equations 1(a-b) are in coupled form. One can decouple them and express
the equation either in terms of spin up or spin down wave functions 
$R_{\pm \frac{1}{2}}$
but the expression loses its transparency. It is thus advisable to use the approach of
Chandrasekhar \cite{ref:c83} by changing the basis and independent variable $r$ to,
$$
r_{*} = r + \frac{2M r_+ + am/\sigma} {r_+ - r_-} {\rm log}
\left({r \over r_+} - 1\right) - \frac{2M r_- + am/\sigma} 
{r_+ - r_-} {\rm log} \left({r \over r_{-}} - 1\right)
\hskip0.2cm ( r > r_{+}).
\eqno{(11)}
$$
where,
$$
{d \over dr_{*}} = {\Delta \over \omega^{2}} {d \over dr}; \ \ \ \ \ \omega^2 =
r^2 + \alpha^2;\ \ \ \  \ \alpha^2 = a^2 + am/\sigma,
\eqno{(12)}
$$
to transform the set of coupled equations 1(a-b) into two 
independent one dimensional wave equations given by:
$$
\left({d \over dr_{*}} -  i \sigma\right)P_{+ {1 \over 2}} = 
\frac {\Delta^{1 \over 2}} {\omega^2} (\lh - i m_p r) P_{- {1 \over 2}};\hskip0.7cm
\left({d \over dr_{*}} + i \sigma\right)P_{- {1 \over 2}} = 
\frac{\Delta^{1 \over 2} } {\omega^2} (\lh + i m_p r) P_{+ {1 \over 2}} .
\eqno{(13)}
$$
Here, ${\cal D}_{0} = {\omega^{2} \over \Delta} ({d \over dr_*} + i \sigma)$
and ${\cal D}^{\dagger}_{0} ={\omega^2 \over \Delta} ({d \over dr_{*}} - i\sigma)$
were used and wave functions were redefined as $R_{- {1 \over 2}} = 
P_{- {1 \over 2}}$ and $\Delta^{1 \over 2} R_{+ {1 \over 2}} = P_{+ {1 \over 2}}$.

We are now defining a new variable,
$$
\theta = tan^{-1} (m_p r/{\lh})
\eqno{(14)}
$$
which yields,
$$
\cos \theta = \frac{\lh} {\surd(\lh^2 + m_p^2 r^2)}, \ \ {\rm and}
\ \ \ \sin \theta = \frac{m_p r} {\surd(\lh^2 + m_p^2r^2)}
$$
and
$$
(\lh \pm i m_p r) = exp ({\pm i \theta}) \surd ({\lh}^{2} + m_p^2 r^2),
$$
so the coupled equations take the form,
$$
\left({d \over dr_{*}} - i \sigma \right) P_{+ {1 \over 2}} = \frac{\Delta^{1 \over 2}}
{\omega^2}({\lh}^2 + m_p^{2} r^2)^{1/2}
P_{- {1 \over 2}} exp\left[-i \tan^{-1} \left({{m_p r} \over {\lh}}\right)\right],
\eqno{(15a)}
$$
and
$$
\left({d \over dr_{*}} + i \sigma\right)P_{- {1 \over 2}} = \frac{\Delta^{1 \over 2}}
{\omega^2}({\lh}^{2} + m_p^{2} r^2)^{1/2} P_{+ {1 \over 2}} 
exp\left[ i \tan^{-1} \left({{m_p r} \over {\lh}}\right)\right].
\eqno{(15b)}
$$

Then defining,
$$
P_{+ {1 \over 2}} = \psi_{+ {1 \over 2}}\  exp\left[-{1 \over 2} 
i\  tan^{-1} \left({{m_p r} \over \lh}\right)\right]
\eqno{(16a)}
$$
and
$$
P_{- {1 \over 2}} = \psi_{- {1 \over 2}}\  exp\left[+{1 \over 2} i 
\ tan^{-1} \left({{m_p r} \over \lh}\right)\right],
\eqno{(16b)}
$$
we obtain,
$$
{{d{\psi}_{+ {1 \over 2}}} \over dr_{*}} - i \sigma 
\left(1 + {\Delta \over \omega^2}{{\lh m_p} \over 2 \sigma}
{1 \over {\lh^2 + m_p^2 r^2}}\right){\psi}_{+ {1 \over 2}} =
{\Delta^{1 \over 2} \over \omega^2}(\lh^2 + m_p^2 r^2)^{1/2}
\psi_{-{1 \over 2}}
\eqno{(17a)}
$$
and
$$
{{d{\psi}_{- {1 \over 2}}} \over dr_{*}} + i \sigma 
\left(1 + {\Delta \over \omega^2}{{\lh m_p} \over 2\sigma}
{1 \over {\lh^2 + m_p^2 r^2}}\right){\psi}_{- {1 \over 2}} =
{\Delta^{1 \over 2} \over \omega^2}(\lh^2 + m_p^2 r^2)^{1/2} \psi_{+{1 \over 2}}.
\eqno{(17b)}
$$

Further choosing $\hat{r}_* = r_{*} + {1 \over 2\sigma} 
{\rm tan}^{-1}({m_p r \over \lh})$ so that
$d{\hat r}_* = (1 + {\Delta \over \omega^2} {{\lh m_p} 
\over 2 \sigma} {1 \over {\lh^2 + m_p^2 r^2}})dr_*$,
the above equations become,
$$
\left(\frac{d} {d{\hat r}_*} - i \sigma\right){{\psi}_{+ {1 \over 2}}} 
= W {{\psi}_{- {1 \over 2}}},
\eqno{(18a)}
$$
and
$$
\left(\frac {d} {d{\hat r}_*} + i \sigma\right){{\psi}_{- {1 \over 2}}} 
= W {{\psi}_{+ {1 \over 2}}}.
\eqno{(18b)}
$$
where,
$$
W = \frac{\Delta^{1 \over 2} (\lh^{2} + m_p^2 r^2)^{3/2} } 
{ \omega^2 (\lh^2 + m_p^2 r^2) + \lh m_p \Delta/2\sigma}.
\eqno{(19)}
$$

Now letting $Z_{\pm} = \psi_{+ {1 \over 2}} \pm \psi_{-{1 \over 2}}$
we can combine the differential equations to give,

$$
\left(\frac{d} {d{\hat r}_*} - W\right) Z_+ = i \sigma Z_- ,
\eqno{(20a)}
$$

and

$$
\left(\frac{d} {d{\hat r}_*} + W\right) Z_- = i \sigma Z_+ .
\eqno{(20b)}
$$

From these equations, we readily obtain a pair of independent 
one-dimensional wave equations,

$$
\left(\frac{d^2} {{d {\hat r}_*}^2} + \sigma^2\right) Z_\pm = V_\pm Z_\pm .
\eqno{(21)}
$$
where, $V_{\pm} = W^{2} \pm {dW \over d\hat{r}_{*}}$,
$$
={{\Delta^{1 \over 2}(\lh^{2} + m_p^{2} r^{2})^{3/2}} \over 
{[ \omega^{2}(\lh^{2} + m_p^{2}
r^{2}) + \lh m_p \Delta/2 \sigma]^{2}}}[\Delta^{1 \over 2}
(\lh^{2} + m_p^{2} r^{2})^{3/2} \pm ((r-M)(\lh^{2} + m_p^{2} r^{2}) 
+ 3m_p^{2} r \Delta)]
$$

$$
\mp {{\Delta^{3 \over 2}(\lh^{2} + m_p^{2} r^{2})^{5/2}} \over 
{[ \omega^{2}(\lh^{2} + m_p^{2} r^{2}) + \lh m_p \Delta/2 \sigma]^{3}}}
[2r(\lh^{2} + m_p^{2} r^{2}) + 2 m_p^{2} \omega^{2} r + \lh m_p (r-M)/\sigma] .
\eqno{(22)}
$$
One important point to note: the transformation of spatial co-ordinate $r$ to
$r_{*}$ (and ${\hat{r}}_{*}$) is taken not only for mathematical simplicity
but also for a physical significance.
When $r$ is chosen as the radial co-ordinate, the decoupled equations
for independent waves show diverging behaviour.
However, by transforming those in terms of $r_{*}$ (and ${\hat{r}}_{*}$)
we obtain well
behaved functions. The horizon is shifted from $r=r_+$ to 
${\hat {r}}_{*}= -\infty$ unless
$\sigma \leq \sigma_s=-am/2Mr_+$ (eq. 11). 
In this connection, it is customary to define
$\sigma_c$ where $\alpha^2=0$ (eq. 13). 
Thus, $\sigma_c=-m/a$. If $\sigma \leq \sigma_s$,
the super-radiation is expected \cite{ref:c83}.

\section{Solution Procedure}

The choice of parameters is generally made in such a way that there is
a significant interaction between the particle and the black hole, 
i.e., when the Compton
wavelength of the incoming wave is of the same order as the outer horizon of
the Kerr black hole. Similarly, the frequency of the incoming
particle (or wave) should be of the same order as inverse of light
crossing time of the radius of the black hole. These yield \cite{ref:bmskc98},
$$
m_p \sim \sigma \sim [M + \sqrt(M^2 - a^2)]^{-1}.
\eqno{(23)}
$$
One can easily check from equation (22) that for $r \rightarrow \infty$
(i.e., ${\hat{r}}_* \rightarrow \infty$) $V_\pm \rightarrow m_p^2$. So the
total energy of the physical particle should greater than square of its
rest mass. So if we expand the total parameter space in terms of
frequency of the particle (or wave), $\sigma$ and rest mass
of the particle, $m_p$, it is clear that $50\%$ of total
parameter space where $\sigma<m_p$
is unphysical (In this case, the energy is such that a particle released from a finite
distance cannot go back to infinity after scattering.), and one need not study this region.
Out of the total physical parameter space 
there are two cases of interest: (1) the waves do not `hit' the potential barrier
and (2) the waves do hit the potential barrier.  To sole these potential
problem first, we replace the
potential barrier by a large number of steps as in the step-barrier problem in
quantum mechanics. Fig. 1 shows one such example of the potential barrier \cite{ref:skcbm00}
$V_+$ (Eq. 22) which is drawn for $a=0.5$, $m_p=0.8$ and $\sigma=0.8$. In reality
we use tens of thousands of steps with suitable variable widths so that the
steps become indistinguishable from the actual function.
The solution of Eq. 22 at $n$th step can be written as \cite{ref:dev76},
$$
Z_{+,n}  = A_{n} exp [ik_n {\hat r}_{*, n}] + B_{n} exp [-ik_n{\hat r}_{*, n}]
\eqno{(24)}
$$
when energy of the wave is greater than the height of the potential barrier.
The standard junction condition is given as \cite{ref:dev76},
$$
Z_{+, n}=Z_{+, n+1} \ \ \ \ {\rm and} \ \ \ \
\frac{dZ_{+}}{d{\hat r}_*}|_n = \frac{dZ_{+}}{d{\hat r}_*}|_{n+1} .
\eqno{(25)}
$$
\begin {figure}
\vbox{
\vskip 4.0cm
\hskip 2.0cm
\centerline{
\psfig{figure=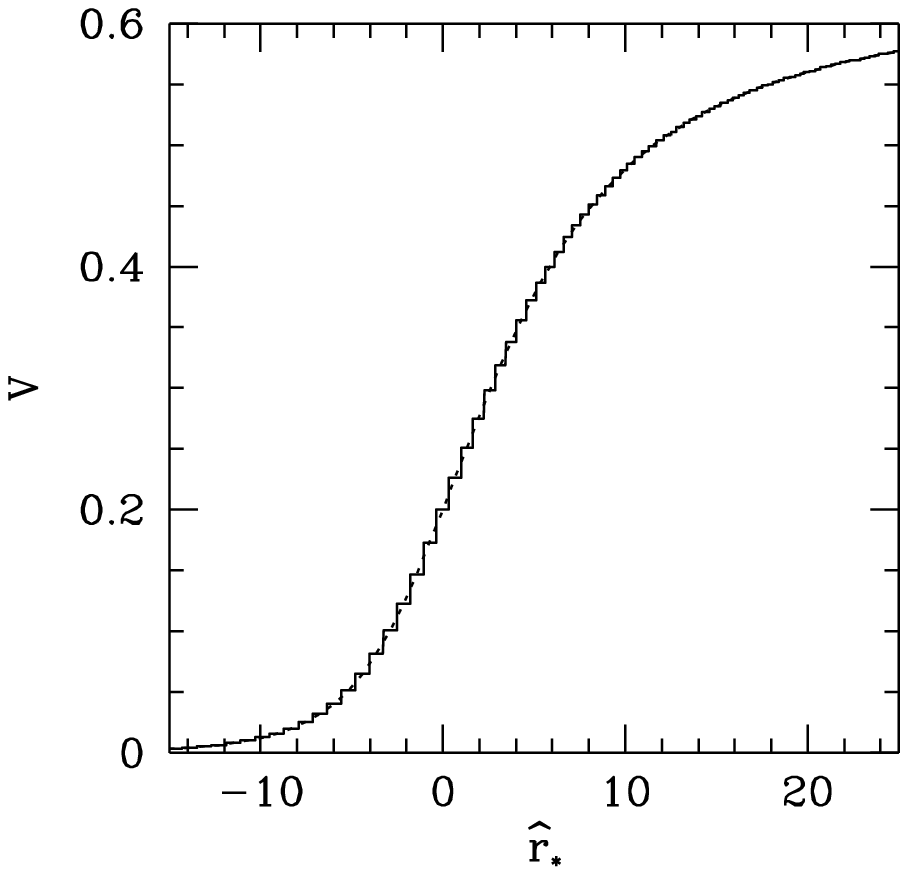,height=10truecm,width=10truecm}}}
\begin{verse}
\vspace{-3.5cm}
\noindent {\small {\bf Fig. 1} : 
Behaviour of $V_+$ (smooth solid curve) for $a=0.5,\ m_p=0.8,\ \sigma=0.8$.
This is approximated as a collection of steps. In reality tens of thousand steps were used
with varying step size which mimic the potential with arbitrary accuracy. }
\end{verse}
\end{figure}

The reflection and transmission co-efficients at $n$th junction are given by:
$$
R_n=  \frac{A_{n+1} (k_{n+1}-k_n)+B_{n+1}(k_{n+1}+k_n)}{A_{n+1}(k_{n+1}+k_n)+B_{n+1}(k_{n+1}-k_n)}; \ \ T_n=1-R_n
\eqno{(26)}
$$
At each of the $n$ steps these conditions were used to connect solutions at successive steps.
Here, $k$ is the wave number ($k=\sqrt{\sigma^2-V_\pm}$) of the wave and $k_n$ is
its value at $n$th step. We use the  `no-reflection' inner boundary condition:
$R \rightarrow 0$ at ${\hat r}_* \rightarrow -\infty $.

For the cases where waves hit on the potential barrier, inside the barrier
(where $\sigma^2 < V_+$) we use the wave function of the form
$$
Z_{+,n}  = A_{n} exp [-\alpha_n {\hat r}_{*, n}] + B_{n} exp [\alpha_n{\hat r}_{*, n}]
\eqno{(27)}
$$
where, $\alpha_n=\sqrt{V_\pm-\sigma^2}$, as in usual quantum mechanics.

\begin {figure}
\vbox{
\vskip 0.0cm
\hskip 3.0cm
\centerline{
\psfig{figure=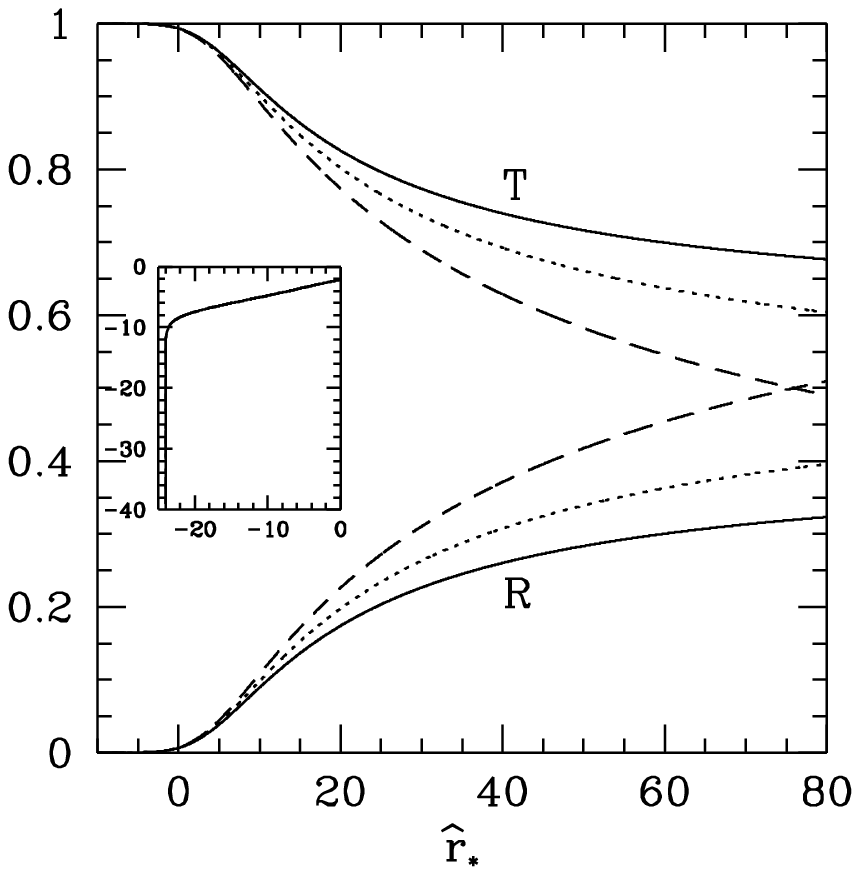,height=10truecm,width=10truecm}}}
\begin{verse}
\vspace{2.0cm}
\noindent {\small {\bf Fig. 2a}:
Reflection ($R$) and transmission ($T$) coefficients of waves
with varying mass as functions of ${\hat r}_*$. $m_p=0.78$ (solid),
$m_p=0.79$ (dotted) and $m_p=0.80$ (long-dashed) are used.
Other parameters are $a=0.5$ and $\sigma=0.8$. Inset shows
$R$ in logarithmic scale which falls off exponentially just outside the horizon.}
\end{verse}
\end{figure}

\begin {figure}
\vbox{
\vskip 5.0cm
\hskip 2.0cm
\centerline{
\psfig{figure=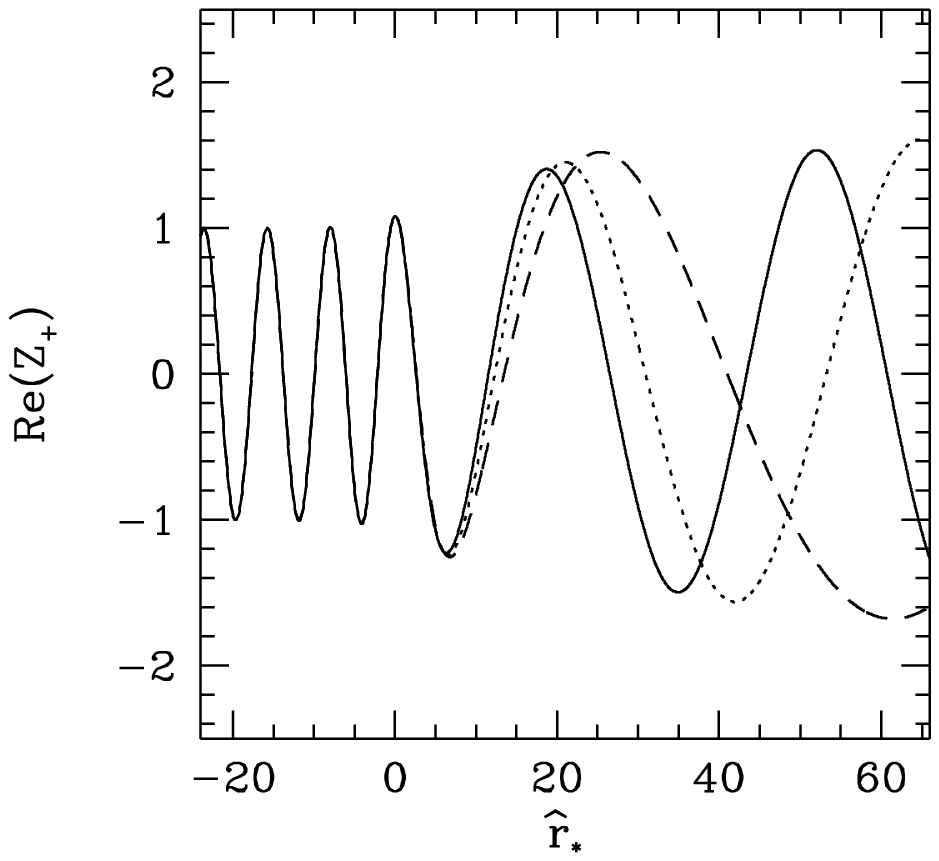,height=10truecm,width=10truecm}}}
\begin{verse}
\vspace{-4.0cm}
\noindent {\small {\bf Fig. 2b}:
Amplitude of Re($Z_+$) of waves
with varying mass as functions of ${\hat r}_*$. $m_p=0.78$ (solid),
$m_p=0.79$ (dotted) and $m_p=0.80$ (long-dashed) are used.
Other parameters are $a=0.5$ and $\sigma=0.8$. }
\end{verse}
\end{figure}

\section{Examples of Solutions}

Fig. 2a shows three solutions [amplitudes of Re($Z_+$)] for parameters: $a=0.5$, $\sigma=0.8$ and
$m_p=0.78,\ 0.79, $ and $0.80$ respectively in solid, dotted and long-dashed curves.
The energy $\sigma^2$ is always higher compared to the height of the potential barrier (Fig. 1)
and therefore the particles do not `hit' the barrier. $k$ goes up and therefore the wavelength
goes down monotonically as the wave approaches a black hole. It is to be noted that
though ours is apparently a `crude' method, it has flexibility and is capable of presenting
insight into the problem, surpassing any other method such as ODE solver packages.
This is because one can choose (a) variable steps depending on steepness of the potential to
ensure uniform accuracy, and at the same time (b) virtually infinite number of steps to follow the
potential as closely as possible. For instance, in the inset, we show $R$ in logarithmic scale
very close to the horizon. All the three curves merge, indicating that the
solutions are independent of the mass of the particle and a closer inspection shows that here, the slope
of the curve depends only on $\sigma$. The exponential dependence of $R_n$ close to the horizon
becomes obvious. Asymptotically, $V_\pm=m_p^2$
(eq. 22), thus, as $m_p$ goes down, the wavelength goes down. In Fig. 2b, we present the
instantaneous values of the reflection $R$ and transmission $T$ coefficients (i.e., $R_n$ and $T_n$ of Eq. 26)
for the same three cases. As the particle mass is decreased, $k$ goes up and  corresponding $R$
goes down consistent with the limit that as $k \rightarrow \infty$, there would be no reflection
at all as in a quantum mechanical problem.

\begin {figure}
\vbox{
\vskip 0.0cm
\hskip 2.0cm
\centerline{
\psfig{figure=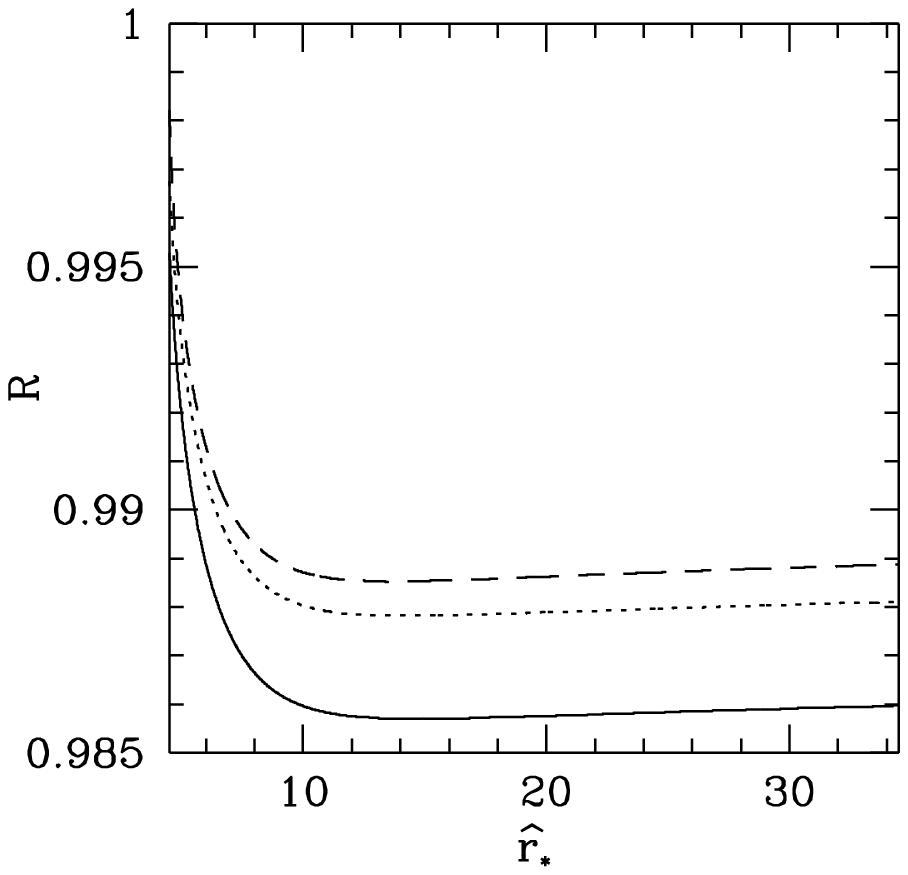,height=10truecm,width=10truecm}}}
\begin{verse}
\vspace{2.0cm}
\noindent {\small {\bf Fig. 3a}}: Reflection ($R$) coefficient of waves
with varying mass as functions of ${\hat r}_*$. $m_p=0.16$ (solid), $m_p=0.164$ (dotted)
and $m_p=0.168$ (long-dashed) are used. 
Other parameters are $a=0.95$ and $\sigma=0.168$.
\end{verse}
\end{figure}

\begin {figure}
\vbox{
\vskip 0.0cm
\hskip 2.0cm
\centerline{
\psfig{figure=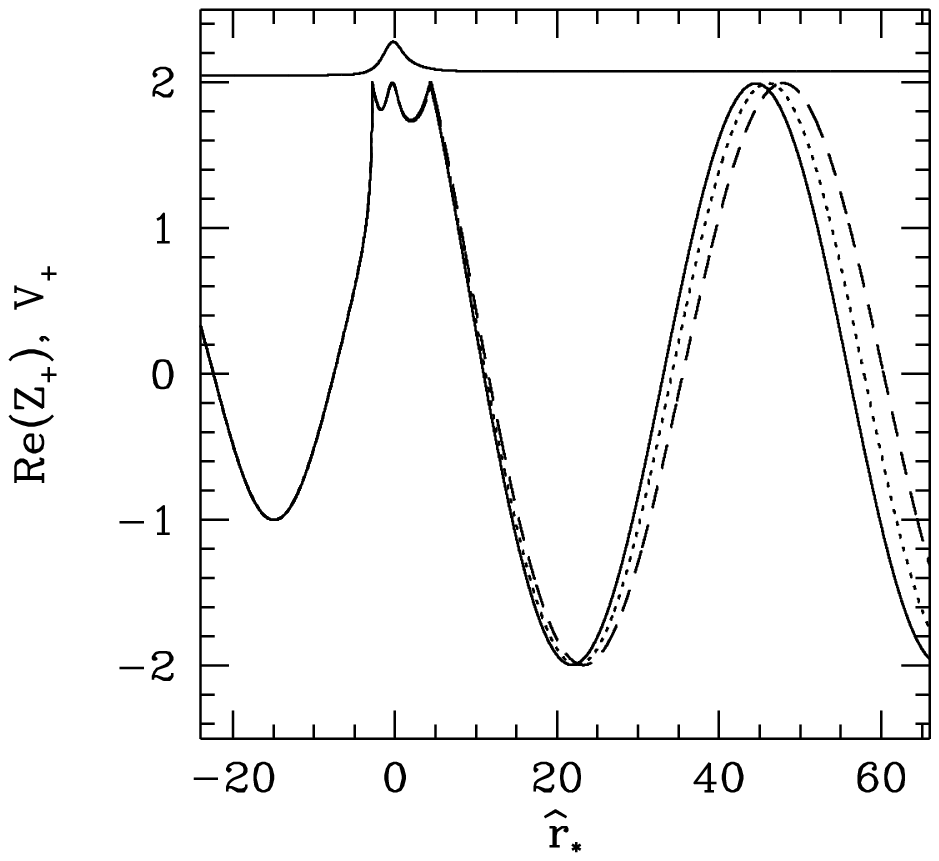,height=10truecm,width=10truecm}}}
\begin{verse}
\vspace{2.0cm}
\noindent {\small {\bf Fig. 3b}}: Amplitude of Re($Z_+$) of waves
with varying mass as functions of ${\hat r}_*$. $m_p=0.16$ (solid), $m_p=0.164$ (dotted)
and $m_p=0.168$ (long-dashed) are used. Nature of potential with $m_p=0.168$ is drawn shifting
vertically by 2.05 unit for clarity. Other parameters are $a=0.95$ and $\sigma=0.168$.
\end{verse}
\end{figure}

Figs. 3(a-b) compare a few solutions where the incoming particles
`hit' the potential barrier. We choose, $a=0.95$, $\sigma=0.168$ and mass of the particle
$m_p=0.16,\ 0.164, \ 0.168$ respectively in solid, dotted and long-dashed curves.
Inside the barrier, the wave decays before coming back to a sinusoidal behaviour,
before entering into a black hole. In Fig. 3b, we plotted the potential (shifted by
2.05 along vertical axis for clarity).  Here too, the reflection coefficient
goes down as $k$ goes up consistent with the classical result that as the
barrier height goes up more and more, reflection is taking place strongly. Note however,
that the reflection is close to a hundred percent. Tunneling causes only a
few percent to be lost into the black hole.

\begin {figure}
\vbox{
\vskip 0.0cm
\hskip -2.0cm
\centerline{
\psfig{figure=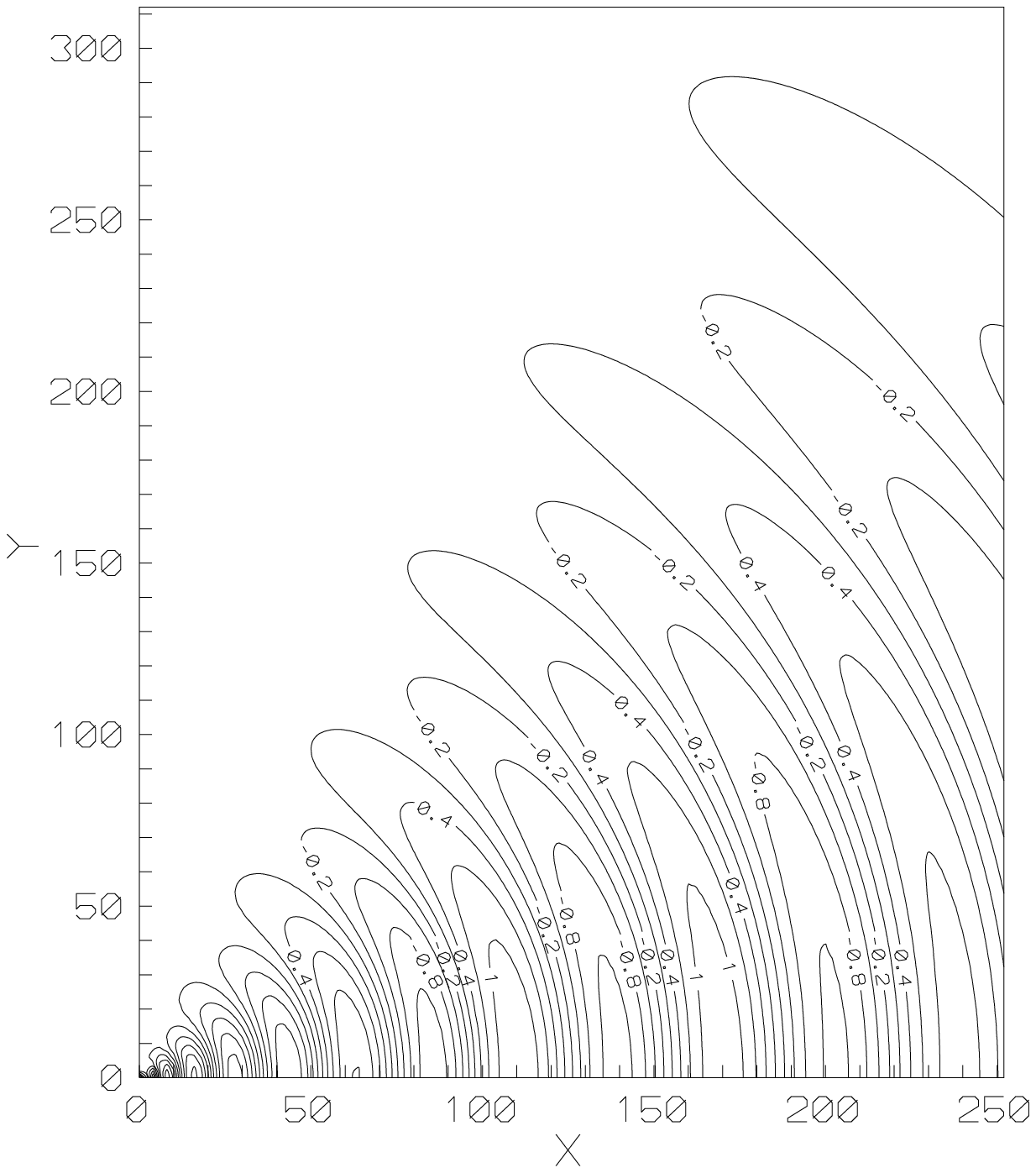,height=14truecm,width=14truecm}}}
\begin{verse}
\vspace{-3.5cm}
\noindent {\small {\bf Fig. 4a}: Contours of constant amplitude 
are plotted in the meridional plane around a black hole. Radial direction on equatorial plane
is  along $X$ axis and the vertical direction and along $Y$.
Both radial and theta solutions have been combined. Parameters
are $a=0.5$, $m_p=0.8$ and $\sigma=0.8$.}
\end{verse}
\end{figure}

\begin {figure}
\vbox{
\vskip -6.0cm
\hskip 0.0cm
\centerline{
\psfig{figure=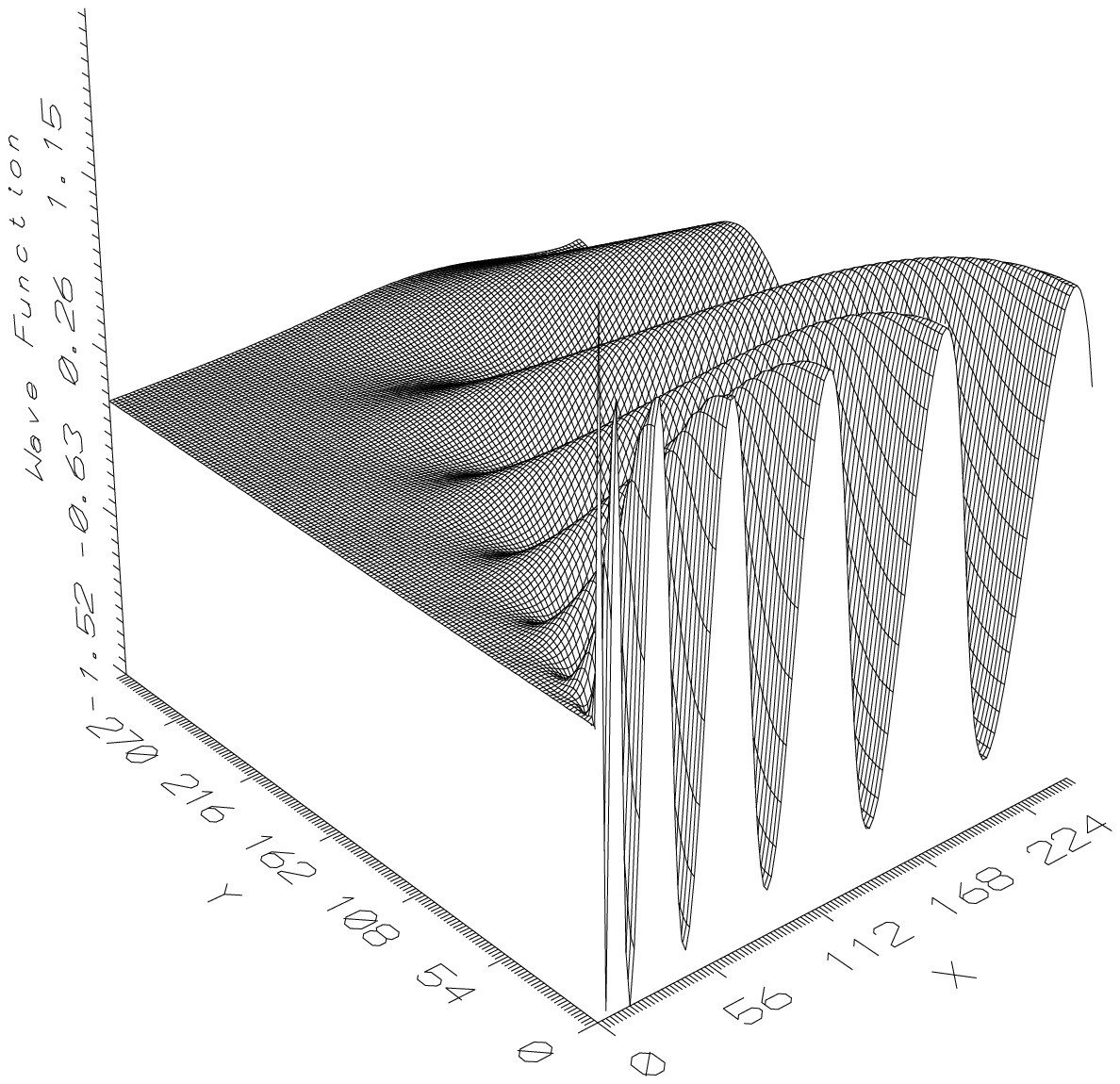,height=14truecm,width=14truecm}}}
\begin{verse}
\vspace{-1.0cm}
\noindent {\small {\bf Fig. 4b}: Three dimensional view of $R_{-1/2}S_{-1/2}$ are plotted
in the meridional plane around a black hole. 
Both radial and theta solutions have been combined. Parameters
are $a=0.5$, $m_p=0.8$ and $\sigma=0.8$.}
\end{verse}
\end{figure}

Figs. 4(a-b) show the nature of the complete wave function
when both the radial and the angular solutions \cite{ref:c84} are included.
Fig. 4a shows contours of constant amplitude of the
wave ($R_{-1/2} S_{-1/2}$) in the meridional plane -- $X$ is along
radial direction in the equatorial plane  and $Y$ is along the vertical direction.
The parameters are $a=0.5$, $m_p=0.8$ and $\sigma=0.8$.
Some levels are marked. Two successive contours have amplitude
difference of $0.1$. In Fig. 4b a three-dimensional nature of
the complete solution is given. Both of these figures clearly
show how the wavelength varies with distance. Amplitude of the spherical
wave coming from a large distance also gets weaker along the
vertical axis and the wave is forced to fall generally along the equatorial
plane, possibly due to the dragging of the inertial frame.

\section{Conclusion}

We review here the scattering of massive, spin-half particles from a
spinning black hole with particular emphasis to the
nature of the radial wave functions and reflection and
transmission coefficients. Here we presented a well known 
quantum mechanical step-potential approach \cite{ref:skcbm00}
but one can verify by any numerical technique that the solution would remain the same.
A modified WKB approximation \cite{ref:bmskc98, ref:bmskc99, ref:bm00} also yields similar
result in Kerr geometry \cite{ref:bmskc00}. The approach presented here
(i.e., step potential approach) is very transparent  since a  complex problem of barrier penetration
in a spacetime around a spinning black hole could be tackled very easily. We report
a few significant observations of these papers that the wave function and $R$, and $T$
behave similarly close to the horizon independent of the initial
parameter, such as the particle mass $m_p$. Particles of different
mass scatter off to a large distance completely differently, thus giving an impression that
a black hole could be treated as a mass spectrograph! When the energy
of the particle becomes higher compared to the rest mass, the reflection
coefficient diminishes as it should it. Similar to a barrier
penetration problem, the reflection coefficient becomes close to a
hundred percent when the wave hits the potential barrier.
Another significant observation is that the reflection and transmission
coefficients are functions of the radial coordinates. This is
understood easily because of the very nature of the potential barrier
which is strongly space dependent which we approximate as a collection
of steps. Combining with the solution of theta-equation, we
find that the wave-amplitude vanishes close to the vertical axis,
possibly due to the frame-dragging effects.

\end{document}